# Effect of grain refinement on enhancing critical current density and upper critical field in undoped MgB$_2$ *ex-situ* tapes

A. Malagoli[1], V. Braccini[1], M. Tropeano[1], M. Vignolo[1], C. Bernini[1], C. Fanciulli[1], G. Romano[1], M. Putti[1], C. Ferdeghini[1], E. Mossang[2], A. Polyanskii[3], D.C. Larbalestier[3]

[1] *CNR-INFM LAMIA, C.so Perrone 24, I-16152 Genova, Italy*
[2] *Grenoble High Magnetic Field Laboratory, C.N.R.S., 25, Avenue des Martyrs, B.P. 166, 38 042 Grenoble Cedex 9, France*
[3] *National High Magnetic Field Laboratory, Florida State University, Tallahassee, FL, 32310*

*Ex-situ* Powder-In-Tube MgB$_2$ tapes prepared with ball-milled, undoped powders showed a strong enhancement of the irreversibility field $H^*$, the upper critical field $H_{c2}$ and the critical current density $J_c(H)$ together with the suppression of the anisotropy of all of these quantities. $J_c$ reached $10^4$ A/cm$^2$ at 4.2 K and 10 T, with an irreversibility field of about 14 T at 4.2 K, and $H_{c2}$ of 9 T at 25 K, high values for not-doped MgB$_2$. The enhanced $J_c$ and $H^*$ values are associated with significant grain refinement produced by milling of the MgB$_2$ powder, which enhances grain boundary pinning, although at the same time also reducing the connectivity from about 12% to 8%. Although enhanced pinning and diminished connectivity are in opposition, the overall influence of ball milling on $J_c$ is positive because the increased density of grains with a size comparable with the mean free path produces strong electron scattering that substantially increases $H_{c2}$, especially $H_{c2}$ perpendicular to the Mg and B planes.

**Introduction**

It is now widely believed that Magnesium Diboride ($MgB_2$) is a potential alternative to the Nb-based superconductors due to its lower raw material costs, its possibility for cryocooler use at around 20 K and its reasonable upper critical field ($H_{c2}$) values. These advantages would be strengthened by further raising the irreversibility field $H^*$ which corresponds approximately to $H_{c2}^{\perp}$ because of the strong effect that $H^*$ has on the whole $J_c(H,T)$ curve. The highest $H_{c2}$ values were obtained on thin films, either through introducing disorder[1] or after C doping, reaching extrapolated values of 70 T in the direction parallel to the *ab* planes.[2] Many efforts focused on the improvement of the critical fields and the critical current density ($J_c$) on polycrystalline samples, wires or tapes. It has been shown how the measured $J_c$, that is the practical parameter useful for applications, is a complex balance between connectivity, $H_{c2}$ and flux pinning induced by grain boundaries and precipitates.[3] The highest $H_{c2}$ values and better performances of the carrying current capability were obtained after doping with C in different forms, i.e. C, SiC, C nanotubes or carbohydrates. Polycrystalline $MgB_2$ samples doped with double-wall carbon nanotubes reached extrapolated $H_{c2}(0)$ of about 44.5 T;[4] $H_{c2}$(4.2 K) higher than 33 T and irreversibility field $H^*$ of 29 T were obtained on *in-situ* $MgB_2$ wires with SiC additions;[5] a strong vortex pinning and $H_{c2}(0)$ values exceeding 40 T were attained in SiC-doped Fe-sheathed $MgB_2$ in-situ tapes.[3] Besides increasing $H_{c2}$, adding SiC to $MgB_2$ also significantly improves $J_c$ of the *in-situ* tapes at high fields, perhaps by introducing a high density of structural nanodefects and nanoscale precipitates.[6] Combining mechanical alloying on powders and C doping in *in-situ* Powder-In-Tube (PIT) tapes allowed to reach $J_c$ of $10^4$ A/cm$^2$ at 14.3 T and 4.2 K.[7]

Often the effect of the doping in increasing the performances in magnetic field is enhanced by a simultaneous reduction in the grain size that increases the grain boundary pinning. A combination of C-substitution-induced $H_{c2}$ enhancement as well as the strong flux pinning centres induced by grain boundaries were indicated as responsible for the high $J_c$–H performances ($10^4$ A/cm$^2$ at 4.2 K, 12 T) in $MgB_2$ tapes prepared by the *in-situ* PIT technique by using maleic anhydride as a dopant.[8] Grain refinement induced by high-energy ball milling C-doped powders was indicated as the reason for the increased $H_{c2}^{\perp}$ and therefore $H^*$ that reached 17.2 T at 4.2 K.[9]

An alternative way of introducing new pinning centres and increasing $H_{c2}$ is neutron irradiation. In an experiment performed on polycrystalline samples $H_{c2}$ exceeded 30 T[10] while at the same time grains boundaries, point-like defects and local variations of the superconducting order parameter contributed to the strong pinning.[11] This method, though scientifically interesting, cannot be applied practically to tapes and wires.

Most of the data reported in the literature in terms of enhanced $H_{c2}$ and $J_c$ has been performed on *in-situ* samples. However, the *ex-situ* process is of great practical importance too. Recently Herrmann *et al.*, reported high $J_c$ values also for undoped *in-situ* tapes, i.e. $10^4$ A/cm$^2$ at 12.1 T and 4.2 K.[7] Very good results have been recently achieved on *ex-situ* samples: we performed ball milling with and without C on *ex-situ* PIT tapes,[12] and reported better $J_c$ values of $10^4$ A/cm$^2$ at 4.2 K and 13 T for the C-doped sample; Senkowicz *et al.* analyzed the effect of ball milling and at the same time of C doping on $H_{c2}$ and $J_c$ in *ex-situ* bulk samples, attaining $10^4$ A/cm$^2$ at 14T.[9] From a performance viewpoint, there thus seems to be no decisive difference between *in-situ* and *ex-situ* MgB$_2$.

In this paper we analyse the behaviour of *ex-situ* PIT tapes prepared with undoped ball-milled MgB$_2$ powders, and report a strong enhancement of both $H_{c2}$ and $J_c(H)$, together with the suppression of the anisotropy of both these quantities. We ascribe such results to strong grain refinement and enhanced grain boundary density, which has the doubly positive effect of enhancing grain boundary vortex pinning and enhancing the electron scattering which increases $H_{c2}$. Our results are thus very important from an applications point of view, since ball milling can be easily applied to production quantities of MgB$_2$ powders, and the *ex-situ* PIT method has been already successfully demonstrated on industrial lengths.[13]

**Experimental Details**

MgB$_2$ powders were prepared from commercial amorphous B (95-97% purity) and Mg (99%

purity): the powders were mixed and underwent a heat treatment at 900°C in Ar. High energy ball milling was conducted in a planetary ball mill with WC jar and media for different times up to 256 hours.

Monofilamentary tapes were fabricated following the *ex-situ* conventional PIT method.[14] $MgB_2$ powders were packed inside Ni or Fe tubes, which were groove rolled and drawn down to a diameter of 2 mm, then cold rolled in several steps to tapes of about 0.35 mm in thickness and 4 mm in width. The superconducting transverse cross section of the conductor was about 0.2 mm$^2$. The conductors were then subjected to a heat treatment at a temperature between 900 and 950 °C for 18 minutes in flowing Argon atmosphere. During the final heat treatment a $Ni_{2.5}MgB_2$ reaction layer forms between the Ni sheath and the $MgB_2$ core,[15] and such a layer does not allow the Ni sheath to be peeled off. Although Fe-sheathed tapes also produce a tiny reaction layer between Fe and B, it is easy to mechanically remove the sheath.[6] Microstructural analysis and resistive measurements of the $MgB_2$ core itself therefore became possible by using a removable Fe sheath, avoiding any undesirable effect or superimposed signal due to the magnetic sheath.

The microstructure of $MgB_2$ was investigated using Scanning Electron Microscopy (SEM) to evaluate the particle size, and X-ray diffraction (XRD) was performed in order to identify the phase, give an estimation of its orientation and calculate the grain size.

Magneto-optical imaging (MOI) was employed to assess the local nature of current flow.[16,17] 3-5 mm long tapes were placed on the cooling finger of a continuous flow optical cryostat located on the X-Y stage of a polarized optical microscope in reflective mode. The Bi-doped garnet indicator film with in-plane magnetization was placed directly on the bare core of the $MgB_2$ tapes to register the normal component of magnetic flux at the tape surface.[16,17]

Short pieces of about 6 mm in length cut from the Ni-sheathed tapes were employed for the magnetization measurements in a 5.5 T MPMS Quantum Design Squid magnetometer. The magnetic field was applied perpendicular to the tape surface and critical current density values were extracted from the *M-H* loops applying the Bean model. Demagnetization corrections are negligible at fields above 0.7 T, that is the saturation field of Ni.[18]

Transport critical current ($I_c$) measurements were performed over ~10 cm long samples at the

Grenoble High Magnetic Field Laboratory (GHMFL) at 4.2 K in magnetic field up to 13 T, applied both perpendicular and parallel to the tape surface, while the current was applied perpendicular to the magnetic field. The criterion for the $I_c$ definition was 1 µV/cm.

Resistive $H_{c2}$ measurements were performed on the MgB$_2$ core of the Fe-sheathed tapes in a 9 T PPMS Quantum Design Physical Properties Measurements System using a measuring current of 1 mA with the magnetic field applied both parallel and perpendicular to the tape surface. $H_{c2}$ was taken to occur at 90% of the resistive transition.

**Results and discussion**

High-energy ball milling was performed on MgB$_2$ powders in order to lower their average grain size and improve the performance in magnetic field.[12] Fig. 1 shows the critical current density $J_c$ as a function of the magnetic field at 5 K as extracted from the *M-H* loops with the field perpendicular to the tape surface. We notice that even if $J_c$ is not affected much at lower fields, a substantial improvement is obtained in high fields by increasing the milling time up to 144 hours, while upon further increasing the milling time the in-field behavior worsens.

In order to investigate the dominant pinning mechanisms, we plotted in Fig. 2 the pinning force at 5 K normalized by its maximum value as a function of the normalized magnetic field $H/H_K = H/H^* = h$. The irreversibility field was determined with a linear extrapolation of the Kramer function $F_K = J_c^{0.5} * B^{0.25}$ – see inset of Fig. 2 -[12] and is therefore indicated as the Kramer field $H_K$. The pinning force behavior is well described by the grain boundary pinning model $F_P = h^{0.5}(1-h)^2$ for all milling conditions. $F_p$ rises from 3.8 GN/m$^3$ for the not-milled powders to 4.5 GN/m$^3$ for the tape prepared with powders milled for 144 hours - suggesting that the ball milling process lowers the grain size and increases the number density of grain boundaries, without introducing additional pinning mechanisms on milling. The enhanced $J_c$ in magnetic field and the higher $H^*$ values are therefore probably due to increased perpendicular $H_{c2}$ values, as it will be discussed later.

After investigating the ball milling parameters, we decided to focus on comparison between the tape prepared with not milled powders (sample NM in the following) and the sample with the best performance in field, i.e. that prepared with powders milled for the optimum 144 hours (sample M).

First of all, we performed SEM analysis on the powders. In Fig. 3 the SEM images are shown, of the powders after synthesis and milling (a), and inside the tapes after cold working and sintering (b). It is immediately clear that after milling the powders are much more homogeneous in size and smaller, and that wire fabrication produces a considerable particle refinement. We performed a statistics on about 700 particles taken from several points of the four samples in order to estimate the average diameter of the powders before and after milling, and into the final tapes. The data were fitted with a Gaussian, as reported in the right part of Fig. 3. The mean diameter values are 1.5 µm and 440 nm respectively for the NM and M powders, indicating a significant – even if not so strong as reported by other authors -[9] decrease in the average diameter of the particles after milling. In the tapes, the powders are even smaller, resulting in an average diameter of 400 and 200 nm respectively in the NM and M tape.

Fig. 4 compares MO images taken on the bare core of the Ni-sheathed tapes prepared with NM (upper panel) and M (lower panel) powders. Fig. 4 (a) and (c) are the optical images of the surface of the two samples, in particular of the Ni-sheathed tapes after removal of the sheath from both sides. Images of the fields produced by the induced currents after cooling in an external magnetic field (the FC state) of 120 mT to a temperature of 5.7 K (b) and 6.4 K (d) respectively for the NM and M sample and then reducing the field to zero are very different. This procedure produces a trapped state throughout the whole sample, in which bright contrast areas correspond to regions of trapped magnetic flux with high $J_c$, while darker areas show lower $J_c$ regions where the local trapped flux density is lower. It is clear that tape M made with milled powders has a more uniform and stronger trapped flux area than the not-milled (NM) tape, even though the $MgB_2$ core size is very similar for both tapes.

In order to characterize the in-field behavior at high magnetic field, the $I_c$ of the two samples were measured up to 13 T at GHMFL. In Fig.5 the transport critical current $I_c$ with field both parallel and perpendicular to the tape surface is shown. First of all there is a remarkable increase in the critical current values in magnetic field upon milling: the $J_c$ threshold of $10^4$ A/cm$^2$ that is reached at 4 T in the NM sample is reached at 10 T after milling. A very striking effect is the disappearance of the $J_c$ anisotropy after ball milling, probably due to multiple causes. First of all, flat rolling during fabrication is likely to have a lesser texturing effect when the grains are smaller, as in the case of the milled powders. Furthermore, $J_c$ is determined by the percolative path cross-section which is apparently larger for the smaller-grain, milled sample. Finally, ball milling introduces disorder which is indicated by the decreased $T_c$ (see Table I) and decreased intrinsic $H_{c2}$ anisotropy.[19]

In order to be able to perform structural and superconducting property characterization of the MgB$_2$ core, we used the same powders to fabricate Fe-sheathed tapes from which we mechanically removed the external sheath. The magnetically evaluated $J_c$ by DC SQUID Magnetometer measurements confirmed that the superconducting behavior of the Ni- and Fe-sheathed tapes was the same.

XRD was performed on the MgB$_2$ cores, as reported in Fig. 6 where the results are compared with the MgB$_2$ starting powders. Both the powders and the cores of the tapes show a high degree of randomly oriented MgB$_2$. A small amount of MgO was identified, particularly in the milled powder sample. WC peaks were also detected in the milled powders. The peaks have been normalized to the most intense MgB$_2$ peak: after milling the peaks weaken and broaden, a sign of a decrease in the particle size.

In the inset of Fig. 6 the relative intensity of the *(002)* peak is shown: in the NM sample it is higher with respect to the pristine powders, showing some degree of texturing; after milling it decreases again, indicating a lower or absent texturing, consistent with the disappearance of anisotropy seen in Fig. 5 after milling. We conclude that the smaller the grains, the less effective is flat rolling in inducing texture.

The lower limit for the average crystallite size can be determined from the half width of the diffraction peaks using the Debye-Scherrer formula $D = \alpha \lambda / \beta \cos\theta$, where $D$ is the mean particle size, $\alpha$ is a geometrical factor (= 0.94), $\lambda$ is the X-Ray wavelength (=1.54056 Å), $\beta$ is the half width of the diffraction peak, and $\theta$ the angular position of the diffraction peak. By analyzing the *(002)* peak we obtain D = 27 nm and 18 nm respectively for the NM and M samples, indicating a reduction in the crystallite size by ball milling. These values are much lower though than the mean particle sizes estimated by SEM analysis, indicating that the particles are polycrystalline agglomerates of much smaller crystallites.

Magnetoresistivity vs. temperature was measured on the $MgB_2$ cores of the Fe-sheathed tapes. In Fig. 7 the resistivities are shown as a function of the temperature for the NM and M samples, and the resistive properties are summarized in Table I. In the inset of Fig. 7 the transition is magnified after normalization of its $\rho$ value at 40 K: the onset (90%) of the critical temperature $T_c$ decreases after milling from 38.6 to 37 K and the transition smears, due to the disorder introduced by milling the powders. We also tried to estimate the effective current-carrying area fraction ($A_F$) following the Rowell analysis,[20] where $A_F = \Delta\rho_{ideal} / [\rho(300\ K) - \rho(40\ K)]$ with $\Delta\rho_{ideal} = 7.3\ \mu\Omega$ cm.[21] The resistivity values are quite high in both samples – 55 and 289 $\mu\Omega$ cm respectively, a first indication of reduced connectivity in both samples. After milling, both $\rho(40\ K)$ and $\rho(300\ K)$ increased while at the same time the residual resistivity ratio RRR almost halved. The calculated effective cross-section decreased from 12% to 8%. We believe that these values provide a lower limit for the actual current-carrying fractions, and although quite low, they are comparable with many other polycrystalline samples reported in the literature where such analysis was carried out.[3,9,22] As reported by other authors too, and also found in our case, the high resistivity values and the inferred poor connectivity do not prevent high $J_c$ values.[23,24] If we try to correct the measured $\rho(40\ K)$ values with the effective cross section we find 6.6 and 22.7 $\mu\Omega$ cm respectively for the NM and M sample (see Table I). A linear relationship between $\rho$ and $T_c$ has been reported for several kinds of samples.[10] With our measured critical temperature values, we expect corrected resistivity values below 10 $\mu\Omega$ cm: it is therefore likely that our samples, in particular the milled one, have an additional intrinsic source of high resistivity such as insulating phases at the grain boundaries – e.g. MgO[25] – given the fact that the controlled atmosphere has not been maintained during the whole process.[26] Furthermore, the inconsistency between the low current-carrying areas and the measured $J_c(0)$ values may also due to the fact that the current paths for the normal-state and superconducting current may not be exactly the same.[27]

In Fig. 8 the $H_{c2}$ vs. $T$ diagram is reported for the NM and M samples perpendicular and parallel to the tape surface. We define the 90% point of the small-current resistive transition, which itself is percolative though the weakly textured or untextured grain structure, as the parallel upper critical field $H_{c2}$. Besides decreasing $T_c$, disorder also increases $H_{c2}$, while at the same time suppressing its anisotropy, as well as the critical current anisotropy. The $H_{c2}$ values reached in the milled sample at 25 K are quite high, especially for undoped samples. They are comparable,[4,28,29] or considerably higher[8,22,30-33] than those reported for SiC or C-doped polycrystalline samples. In Fig. 9 we also report the 10% points of the resistive transitions. These provide a marker of the irreversibility line[30] and $H_{c2}$ perpendicular[9], although occurring between $H_{c2}^{\perp}$ and $H_{c2}^{//}$, given that our tapes are far from being completely oriented. Fig. 9 compares these evaluations to a critical state evaluation of $H^*$(4.2 K) defined by $J_c = 500$ A/cm$^2$, which corresponds to a critical current of 1 A flowing in the tapes. Such an estimate of $H^*$ is clearly well below that deduced from the 10% points of the resistive transitions, even if they do show the same dependence on milling state. After milling, $H^*$ doubled reaching about 14 T at 4.2 K, comparable with the best undoped *in-situ* samples.[7]

We can now try to draw some conclusions from the competing influence of connectivity, $H_{c2}$ and flux pinning. At all the measured magnetic fields, both from magnetic and transport measurements, the milled sample has a higher $J_c$, consistent with its almost doubled $H_{c2}$. Near self field the critical current densities are very similar, showing that milling does not degrade $J_c$ at all, even though the connectivity of the milled sample is about 1/3 lower (see Table I). There is no change in the pinning mechanisms after milling. The dominant grain boundary pinning just rises from 3.8 to 4.5 GN/m$^3$ due to the increased grain boundary density. $J_c$ is in fact proportional to the inverse grain size.[9,34] We see that particle refinement by ball milling decreases the crystallite sizes as well, increasing their number density. We think that the smaller crystallites are responsible for the raised $H_{c2}$ by scattering electrons, because of the mean free path reduction enhances $H_{c2}$, and the resistivity.[10] The resistivity values corrected after calculating the effective cross-section following Rowell[20] (Table I) for the M sample at 40 K is 22.7 $\mu\Omega$ cm. However as noted above this may still be an overestimate since $T_c$ is not strongly suppressed. We believe that the real resistivity is below 10 $\mu\Omega$ cm, which would lead to a mean free path around 10 nm.[10] In fact the crystallite size inferred from the XRD measurements on the milled sample is around 18 nm, which is consistent with our conclusion that the crystallites act as electron scattering centers and are therefore responsible for the substantially increased $H_{c2}$.

In this delicate balance between flux pinning, connectivity and $H_{c2}$, we therefore think that the poor connectivity plays a lesser role while the refined crystallite size seems to be the most significant parameter responsible for enhancing $J_c$ at high magnetic fields, as reported by several authors.[35,36]

**Conclusions**

In this work we studied the effect of ball milling of the precursors $MgB_2$ undoped powders on the connectivity, $H_{c2}$ and $J_c$ of *ex-situ* PIT tapes. We optimized the milling time to obtain the best in – field performance, and reached the $J_c$ value of $10^4$ A/cm$^2$ at 4.2 K and 10 T, an irreversibility field of about 14 T at 4.2 K, and $H_{c2}$ of 9 T at 25 K, all high values for not-doped $MgB_2$. The powder milling process produces a grain refinement; at the same time the calculated connectivity decreases from about 12% to 8% and the anisotropy of both the critical current and the upper critical field disappears. The increased density of grain boundaries increases the pinning force and at the same time substantially increases $H_{c2}$, since the crystallite size is comparable with the mean free path.

The importance of such method of increasing the in-field performances of Magnesium Diboride lies in the fact that the milling of commercial or home-made undoped $MgB_2$ powders is immediately scalable to industrial quantities for the fabrication of long-length multifilamentary *ex-situ* tapes with the same improved performances.

**Acknowledgements**

Financial support by Columbus Superconductors S.p.A., "Compagnia di S. Paolo"and by the "Transnational Access-Specific Support Action" Program - Contract no. RITA-CT-2003-505474 of the European Commission and by the EU FP6 project NMP-3-CT2004-505724 is acknowledged. Work at Florida State University was supported by DOE Office of High Energy Physics and the Focused Research Group on $MgB_2$ DMR-0514592..

|  | $\rho(300K)$ ($\mu\Omega$ cm) | $\rho(40K)$ ($\mu\Omega$ cm) | RRR | $\Delta\rho$ ($\mu\Omega$ cm) | $A_F$ | $\rho(40K)$ corrected ($\mu\Omega$ cm) | $T_c$ onset (K) | $\Delta T_c$ (K) |
|---|---|---|---|---|---|---|---|---|
| Not milled | 116 | 55 | 2.1 | 61 | 12% | 6.6 | 38.6 | 0.4 |
| Milled | 382 | 289 | 1.32 | 93 | 8% | 22.7 | 37.0 | 1.6 |

TABLE I. Resistivity data and critical temperature of the tapes prepared with not milled and 144 hours-milled powders.

FIGURE CAPTIONS

FIG. 1. Magnetic $J_c$ vs. magnetic field at 5 K in perpendicular orientation for tapes fabricated with MgB$_2$ powders milled for different times. $J_c$ was evaluated from SQUID magnetization measurements and are thus lower than conventional transport.

FIG. 2. Behavior of the normalized pinning force vs normalized magnetic field $h = H/H_K$ at T = 5 K as extracted from the magnetic $J_c$ measurements. A comparison with grain boundary pinning model $F_P = h^{0.5} (1-h)^2$ is also reported (continuous line). In the inset, the Kramer function $F_K = J_c^{0.5} * B^{0.25}$ is reported as a function of the field and the linear extrapolation to estimate the irreversibility field H*$_K$ used to plot $F_p$.

FIG. 3. SEM images of the not milled (NM) and 144 hours-milled (M) powders (a) and NM and M tapes (b). Below, the Gaussian fit of particle size distribution is reported after measuring about 500 grains for each sample. The average diameter is 1.4 µm and 440 nm for the NM and M powders (dotted lines), and 400 and 200 nm for the NM and M tapes, respectively.

FIG. 4. MO images of the NM (upper panel) and M (lower panel) tapes after removing the Ni sheath. (a) and (c) are the optical images of the surfaces of the NM and M samples; (b) and (d) the images at H = 0 after FC in H = 120 mT, T = 5.7 K and 6.4 K respectively.

FIG. 5. Transport critical current and corresponding calculated critical current density as measured up to 13 T at GHMFL in He bath for the tapes with not milled and 144 hours-milled powders. The field was both perpendicular and parallel to the tapes surface.

FIG. 6. XRD of MgB$_2$ powders compared with the core of Fe-sheathed tapes fabricated with not milled and 144 hours-milled powders. The indexed peaks come from the MgB$_2$ phase. In the inset, the *(002)* peak is magnified for the three samples.

FIG. 7. Resistivity vs. temperature curves of the MgB$_2$ tapes (after removal of the Fe sheath) prepared with not milled and 144 hours-milled powders. In the inset the superconducting transition is magnified, normalized at 40 K.

FIG. 8. $H_{c2}$ vs. $T$ diagram (i.e. 90% of normal state resistivities from the resistive transitions in magnetic field) for the samples prepared with not milled (NM) and 144 hours-milled (M) powders in both the perpendicular and parallel direction with respect to the tape surface. The disappearance of the anisotropy after milling is clear.

FIG. 9. 10% of normal state resistivities from the resistive transitions in magnetic field for the samples prepared with not milled (NM) and 144 hours-milled (M) powders and measured in both perpendicular and parallel direction. At 4.2 K an estimate of the irreversibility field as the point at which the transport $J_c$ is low enough (500 A/cm$^2$) is reported.

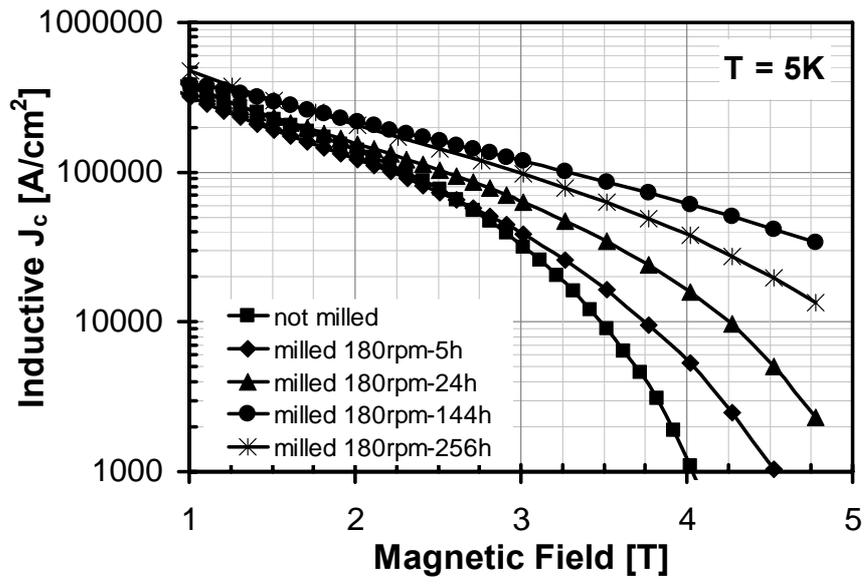

**Figure 1.**

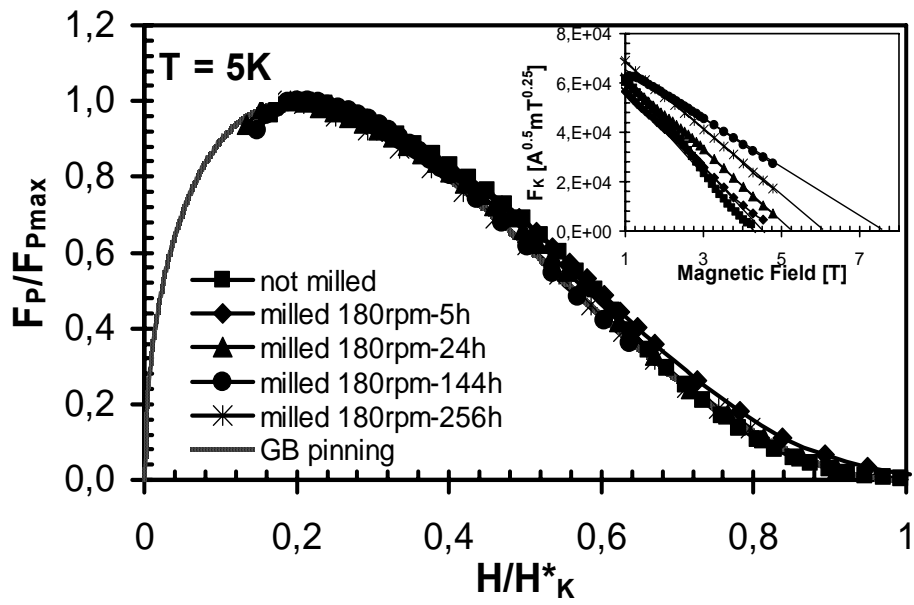

**Figure 2.**

**Figure 3.**

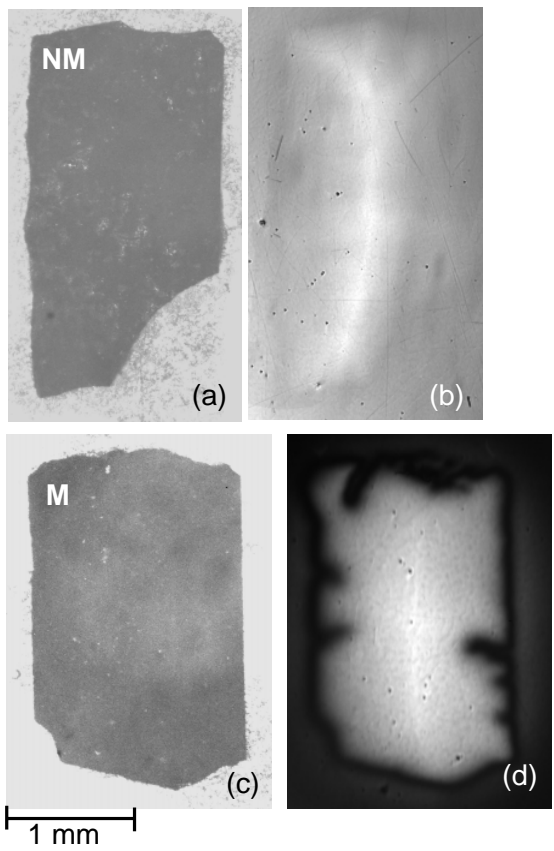

Figure 4

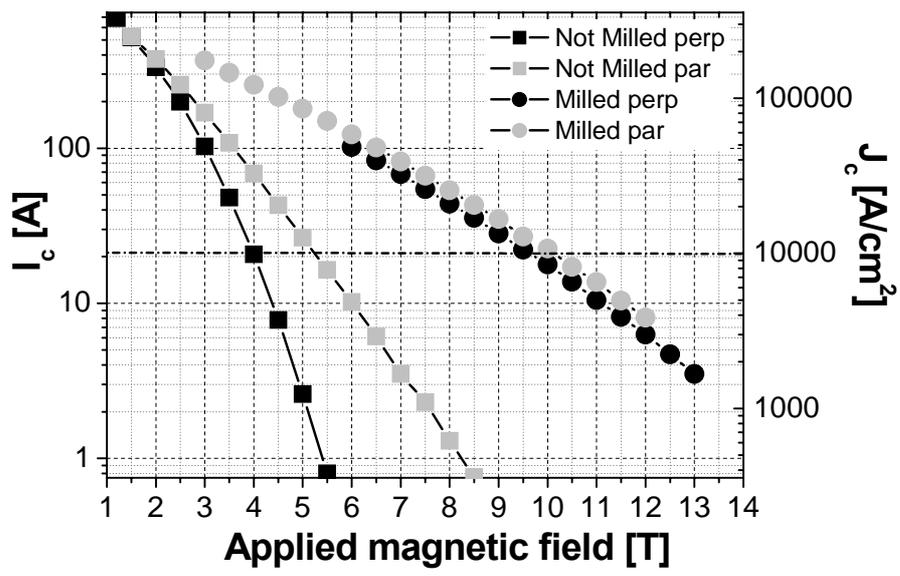

**Figure 5.**

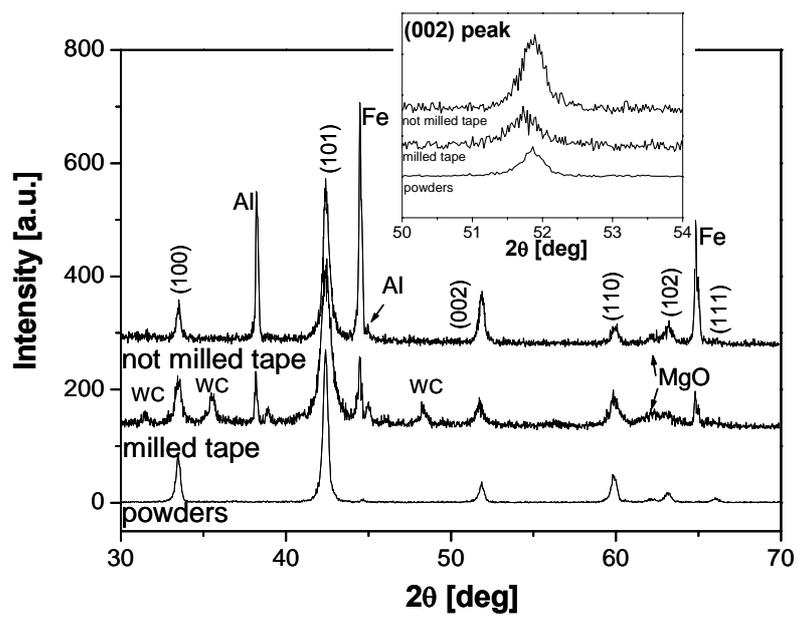

**Figure 6.**

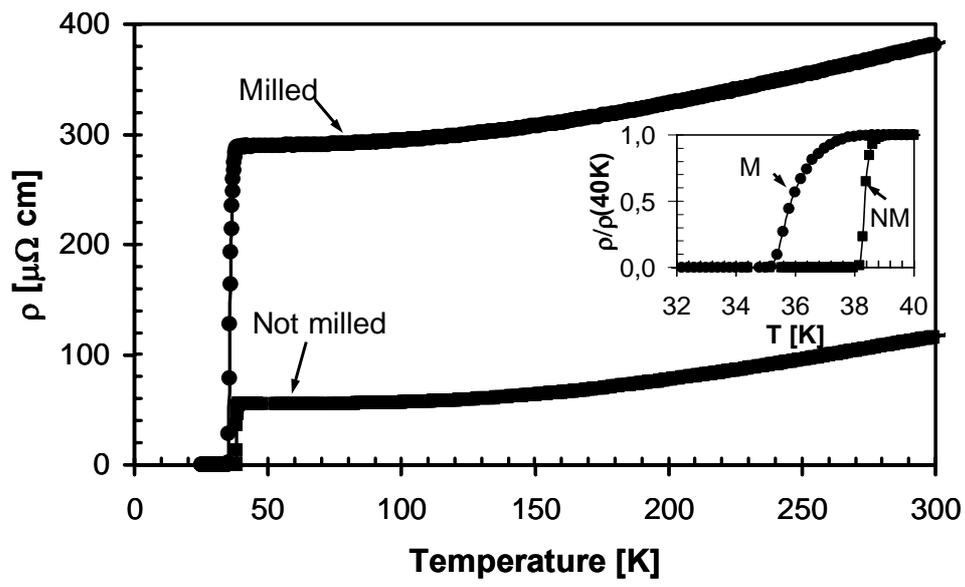

**Figure 7.**

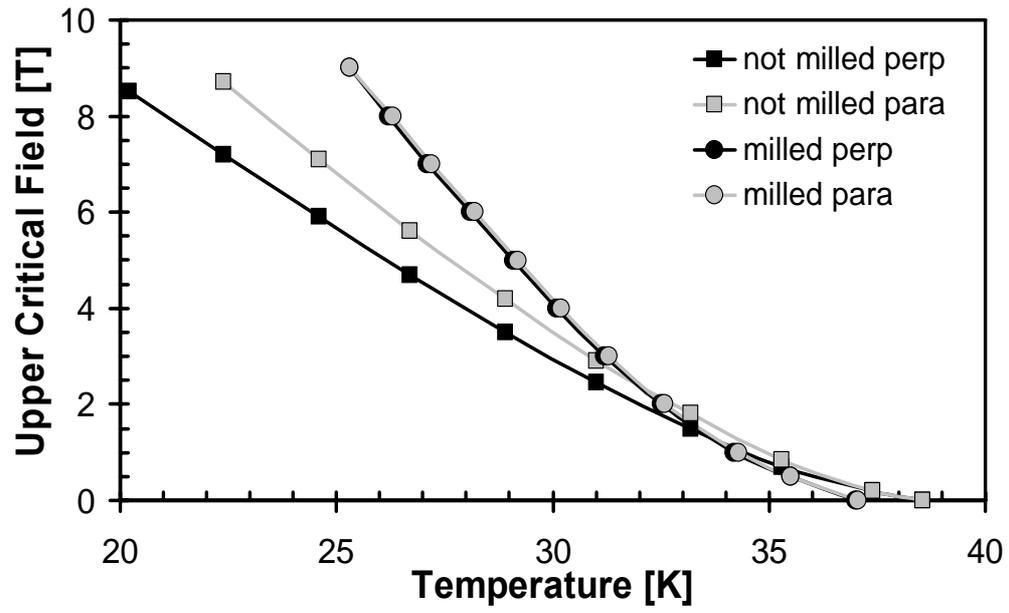

**Figure 8.**

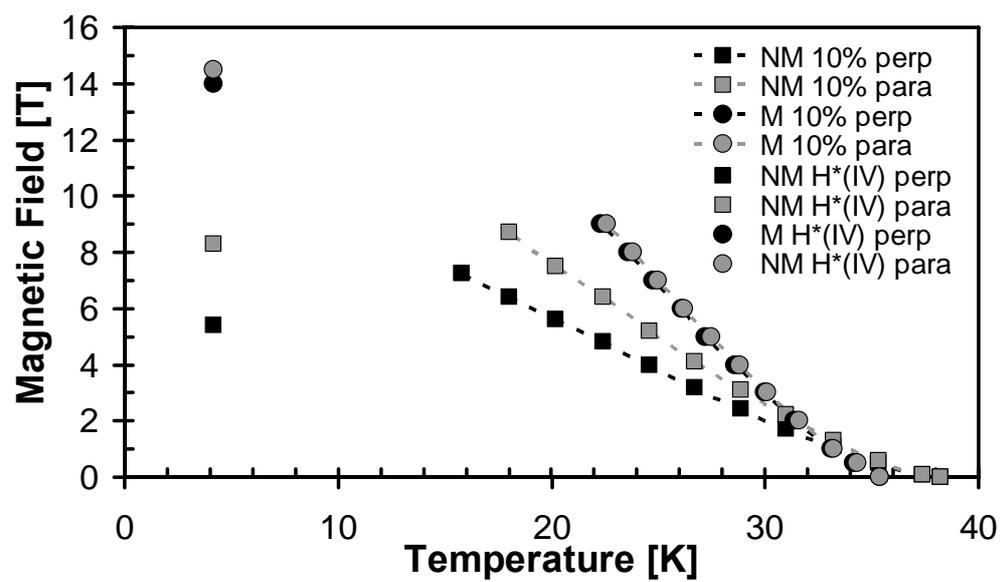

**Figure 9.**